\newcommand{\MSbar}{$\overline{\rm MS}$}
\newcommand{\DRbar}{$\overline{\rm DR}$}
\newcommand{\DRbarprime}{$\overline{\rm DR}'$}
\newcommand{\lnbar}{{\overline{\rm ln}}}
\newcommand{\lnbarx}{\overline{\rm ln}x}
\newcommand{\lnbary}{\overline{\rm ln}y}
\newcommand{\lnbarz}{\overline{\rm ln}z}
\newcommand{\lnbaru}{\overline{\rm ln}u}
\newcommand{\lnbarv}{\overline{\rm ln}v}
\newcommand{\dilog}{{\rm Li}_2}
\newcommand{\trilog}{{\rm Li}_3}
\newcommand{\Tbar}{\overline{T}}

\documentclass[twocolumn,%showpacs,
preprintnumbers,amsmath,amssymb,aps,
floatfix]{revtex4}

\allowdisplaybreaks

\usepackage{axodraw}
\usepackage{graphicx}% Include figure files
\usepackage{bm}% bold math
\begin{document}
\renewcommand{\theequation}{\arabic{section}.\arabic{equation}}

hep-ph/0307101, FERMILAB-Pub-03/202-T

\title{Evaluation of two-loop self-energy basis integrals using 
differential equations}

\author{Stephen P. Martin}
\affiliation{
Physics Department, Northern Illinois University, DeKalb IL 60115 USA\\
{\rm and}
Fermi National Accelerator Laboratory, PO Box 500, Batavia IL 60510}

%\today

%Remember: There Is No Cabal. \today

\begin{abstract}
I study the Feynman integrals needed to compute two-loop self-energy
functions for general masses and external momenta. A convenient basis for
these functions consists of the four integrals obtained at the end of
Tarasov's recurrence relation algorithm. The basis functions are modified
here to include one-loop and two-loop counterterms to render them finite;
this simplifies the presentation of results in practical applications.  I
find the derivatives of these basis functions with respect to all
squared-mass arguments, the renormalization scale, and the external
momentum invariant, and express the results algebraically in terms of the
basis. This allows all necessary two-loop self-energy integrals to be
efficiently computed numerically using the differential equation in the
external momentum invariant. I also use the differential equations method
to derive analytic forms for various special cases, including a
four-propagator integral with three distinct non-zero masses.
\end{abstract}


\maketitle

%\tableofcontents

\section{Introduction\label{intro}}
\setcounter{equation}{0}

The comparison of data with the predictions of the Standard Model, and
candidate extensions of it, requires the kind of accuracy obtained from
two-loop and even higher-order calculations. As a forward-looking example, 
if supersymmetry
proves to be correct then the Large Hadron Collider will be able to
measure the mass of the lightest neutral Higgs scalar boson to an accuracy
of order 100 MeV, and a future linear collider will certainly do
better \cite{Higgsexp}. In contrast, even assuming perfect knowledge of
all input parameters, the present theoretical uncertainty is probably at 
least 10
times larger \cite{Higgstheory}.

The motivation for the present paper is to
facilitate routine calculations of self-energies, and thus pole masses,
for particles in any field theory. A key step in this process
is the evaluation of the necessary two-loop
integrals. It has become clear that analytical methods will only work in
special cases, so practical numerical methods are needed. In this
paper, I will build on the many important advances that have been made in
this area 
\cite{Tarasov:1997kx}-\cite{Mertig:1998vk}, with the goal of streamlining both
computations and presentations of results for self-energies.

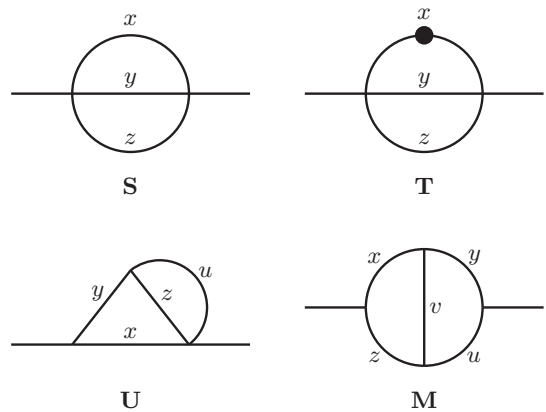
\begin{figure}[b]
\begin{picture}(108,80)(-54,-40)
\SetWidth{0.9}
\Line(-45,0)(-22,0)
\Line(45,0)(22,0)
\CArc(0,0)(22,0,180)
\CArc(0,0)(22,180,360)
\Line(-22,0)(22,0)
\Text(0,28)[]{$x$}
\Text(0,5)[]{$y$}
\Text(0,-17)[]{$z$}
\Text(0,-35)[]{$\bf S$}
\end{picture}
\begin{picture}(108,80)(-54,-40)
\SetWidth{0.9}
\Line(-45,0)(-22,0)
\Line(45,0)(22,0)
\CArc(0,0)(22,0,180)
\CArc(0,0)(22,180,360)
\Line(-22,0)(22,0)
\Vertex(0,22){3.5}
\Text(0,30.2)[]{$x$}
\Text(0,5)[]{$y$}
\Text(0,-17)[]{$z$}
\Text(0,-35)[]{$\bf T$}
\end{picture}
\begin{picture}(108,80)(-54,-40)
\SetWidth{0.9}
\Line(-45,-14)(-22,-14)
\Line(45,-14)(22,-14)
\Line(-22,-14)(0,14)
\Line(22,-14)(0,14)
\CArc(11,0)(17.8045,-51.843,128.157)
\Line(-22,-14)(22,-14)
\Text(0,-9.67)[]{$x$}
\Text(-12.3,5.3)[]{$y$}
\Text(13.8,4.8)[]{$z$}
\Text(28.3,13.4)[]{$u$}
\Text(0,-35)[]{$\bf U$}
\end{picture}
\begin{picture}(108,80)(-54,-40)
\SetWidth{0.9}
\Line(-45,0)(-22,0)
\Line(45,0)(22,0)
\CArc(0,0)(22,0,180)
\CArc(0,0)(22,180,360)
\Line(0,-22)(0,22)
\Text(-18.8,18.8)[]{$x$}
\Text(19,19)[]{$y$}
\Text(-18.8,-18.8)[]{$z$}
\Text(18.8,-18.8)[]{$u$}
\Text(4.5,0)[]{$v$}
\Text(0,-35)[]{$\bf M$}
\end{picture}
\caption{\label{fig:STUM} Feynman diagrams for the two-loop basis 
integrals.}
\end{figure}

Tarasov \cite{Tarasov:1997kx} has provided a solution to the problem of
reducing two-loop self-energy integrals to a minimal basis, such that any
scalar integral can be represented as a linear combination of integrals of
just four types, plus terms quadratic in one-loop integrals. (Other useful
reduction algorithms are presented in \cite{Weiglein:hd} and
\cite{Ghinculov:1997pd}.) Tarasov's algorithm relies on the integration
by parts technique \cite{Tkachov:wb} and
repeated use of recurrence relations involving integrals in different
numbers of dimensions \cite{Tarasov:1996br}. 
The two-loop scalar basis integrals remaining 
after applying this algorithm have the topologies shown in Figure 
\ref{fig:STUM}.
They are the three-propagator ``sunrise" diagram $S$, a diagram $T$ which
is obtained from the sunrise diagram by differentiating with respect to
one of the squared masses, a four-propagator diagram $U$, and the
five-propagator ``master" \cite{Broadhurst:1987ei} diagram $M$.

Consider a generic two-loop integral $F_i(s;x,y,\ldots)$, which 
depends on the external momentum invariant 
\begin{equation}
s = -p^2,
\end{equation}
[using either a Euclidean or a signature ($-$$+$$+$$+$) metric] and
propagator squared masses $x,y,\ldots$. For special values of the
arguments, it may be possible to compute $F_i$ analytically in terms of
polylogarithms\cite{Lewin} or Nielsen's generalized 
polylogarithms\cite{Kolbig}. This requires \cite{Scharfthesis} 
that there is no three-particle cut of the diagram for which the
three cut masses $m_1,m_2,m_3$, the invariant $\hat s$ 
for the total momentum flowing across the cut, and the four quantities
\begin{equation}
\hat s-(m_1\pm m_2 \pm m_3)^2 
\end{equation} 
are all non-zero. 
Many analytical results for various such special cases have been worked 
out \cite{Rosner}-\cite{Martin:2001vx}. There are also expansions 
\cite{Smirnov:rz}-\cite{Berends:1994sa}
in large and small values of the external momentum invariant,
and near the thresholds and pseudo-thresholds 
\cite{Berends:1996gs}-\cite{Groote:2000kz}.
Integral representations 
\cite{Kreimer:1991jv}-\cite{Passarino:2001wv}
allow for systematic numerical evaluations.

In this paper I rely instead on the differential equation method 
\cite{Kotikov:1990kg}-\cite{Caffo:2002wm} for
evaluating the basis integrals. 
The idea is to take advantage 
of the fact that
the basis integrals $F_i$ satisfy a set of coupled
first-order linear ordinary
differential equations in $s$, of the form
\begin{eqnarray}
s \frac{d}{ds} F_i &=& \sum_j C_{ij} F_j + C_i.
\label{genericseq}
\end{eqnarray}
Here $C_{ij}$ and $C_i$ are ratios of polynomials in $s$ and
the squared masses. (If we include only genuine two-loop functions in the
set $F_i$, then $C_i$ will also include terms linear and quadratic in the 
one-loop functions, which are known analytically and present no problems.)
The values of the functions $F_i$ are known analytically
at $s=0$. So one can integrate the differential equations from the 
initial conditions at $s=0$ 
to the desired value of $s$ using
well-known  numerical techniques such as Runge-Kutta.
For the integrals of
the type $S,T,U$, this has already been done and explained in detail in 
\cite{Caffo:1998du}-\cite{Caffo:2002wm}. 
Here, I will extend these results to include the 
master integral $M$, and present results for $S,T,U$ integrals in
a different basis which may be more convenient for some purposes.

In order to find the differential equations in $s$ that the basis
integrals satisfy, I proceed by first calculating the derivatives
of the basis integrals with respect to their propagator squared-mass
arguments. 
Using Tarasov's recurrence relations,
these derivatives are expressed algebraically in terms of the basis 
functions, in the linear form:
\begin{eqnarray}
\frac{\partial}{\partial x} F_i
%(s; x,y,\ldots) 
&=& \sum_j K_{xij} F_j + 
K_{xi}.
\end{eqnarray}
The equations (\ref{genericseq}) in $s$ will then follow by 
elementary
dimensional analysis, using the known dependence of the basis functions
on the renormalization scale. 
The derivatives of the basis functions with respect to the squared masses
are also
useful in their own right, 
since each derivative adds an extra power of the corresponding propagator
in the denominator. This provides a simplified algebraic algorithm 
for computing integrals with arbitrary powers of the 
propagators present in the master integral topology.

The rest of this paper is organized as follows. Section \ref{conventions}
defines the basis integrals, and gives conventions and notations.  
Section \ref{mderivs} presents the derivatives of the basis integrals with
respect to their squared-mass arguments. In section \ref{diffeqs}, I give 
the
differential equations in $s$ satisfied by the basis functions. The
numerical integration of the differential equations near $s=0$ relies on
expansions for small $s$, which are provided in section \ref{exps}. 
Section \ref{analytical} presents some analytic expressions for the basis
functions in special cases that are useful both for comparison with the
literature and for practical purposes. Section \ref{numerical} describes
the numerical computation of the basis integrals, and gives two examples.

\section{Conventions and Setup\label{conventions}}
\setcounter{equation}{0}

The loop functions in this papers are defined by scalar Euclidean 
momentum integrals regularized by 
dimensional reduction to 
$d=4-2\epsilon$ dimensions. Let us define a loop factor
\begin{equation}
C = (16 \pi^2) \frac{\mu^{2\epsilon}}{(2 \pi)^d} 
= (2 \pi \mu)^{2 \epsilon}/\pi^2 .
\end{equation}
The regularization scale $\mu$ is related to the renormalization scale 
$Q$ (in the \MSbar~scheme \cite{msbar}, or the \DRbar~scheme 
\cite{drbar} for supersymmetric
theories, or in the
\DRbarprime~scheme \cite{drbarprime} for softly broken supersymmetric 
theories) by
\begin{eqnarray}
Q^2 &=& 4 \pi e^{-\gamma} \mu^2 .
\end{eqnarray}
Logarithms of dimensionful quantities are always given in terms of
\begin{eqnarray}
\lnbar X &\equiv& {\rm ln}(X/Q^2) .
\end{eqnarray}
The loop integrals are functions of a common external momentum
invariant $s$ as explained in the Introduction. (Note that the
sign convention is such that
for a stable physical particle with mass $m$, there is a pole at 
$s = m^2$.) Throughout this paper, $s$ should be taken to have an
infinitesimal positive imaginary part. Since all functions in any given
equation
have the same $s$, it will not be included explicitly
in the list of arguments.

The one-loop self-energy integrals 
\cite{Passarino:1978jh}
are defined as:
\begin{eqnarray}
{\bf A}(x) &=& 
C \int d^d k \frac{1}{[k^2 +x]}  ,
\label{defineboldA}
\\
{\bf B}(x,y) &=&
C \int d^d k   \frac{1}{[k^2 +x] [(k-p)^2 +y]}
.
\label{defineboldB}
\end{eqnarray}
The two-loop integrals are defined as:
\begin{widetext}
\begin{eqnarray}
{\bf S}(x,y,z) &=& 
C^2\int d^d k \int d^d q  
\frac{1}{[k^2 +x] [q^2 +y] [(k+q-p)^2 +z]} ,
\label{defineboldS}
\\
{\bf T}(x,y,z) &=& 
%C^2\int d^d k \int d^d q
%\frac{1}{[k^2 +x]^2 [q^2 +y] [(k+q-p)^2 +z]} =
-\frac{\partial}{\partial x} {\bf S}(x,y,z) ,
\label{defineboldT}
\\
{\bf U}(x,y,z,u) &=&
C^2 \int d^d k \int d^d q  
\frac{1}{[k^2 +x] [(k-p)^2 +y] [q^2 +z] [(q+k-p)^2 + u]} ,
\label{defineboldU}
\\
{\bf M}(x,y,z,u,v) &=& C^2 \int d^d k \int d^d q  
\frac{1}{[k^2 +x][q^2 + y] [(k-p)^2 +z][(q-p)^2 +u][(k-q)^2 +v]} .
\label{defineboldM}
\end{eqnarray}
I find it convenient to introduce modified integrals in which appropriate
divergent parts have been 
subtracted.
At one-loop order, define the finite and $\epsilon$-independent integrals:
\begin{eqnarray}
A(x) &=& \lim_{\epsilon \rightarrow 0} \left [{\bf A}(x) + x/\epsilon \right ]
= x (\lnbarx-1) ,
\\
B(x,y) &=& 
\lim_{\epsilon \rightarrow 0} \left [ {\bf B}(x,y) - 1/\epsilon \right ] 
= -\int_0^1 dt \>\lnbar [t x + (1-t) y -t (1-t) s ]
.
\end{eqnarray}
At two loops, let 
\begin{eqnarray}
S(x,y,z) &=&
\lim_{\epsilon \rightarrow 0} \left [
{\bf S}(x,y,z) 
- S^{(1)}_{\rm div} (x,y,z) - S^{(2)}_{\rm div} (x,y,z)
\right ] ,
\end{eqnarray}
where 
\begin{eqnarray}
S^{(1)}_{\rm div} (x,y,z) &=& 
\left ({\bf A}(x) + {\bf A}(y) + {\bf A}(z) \right )/\epsilon ,
\\
S^{(2)}_{\rm div} (x,y,z) &=& 
(x+y+z)/2\epsilon^2 + (s/2-x-y-z)/2 \epsilon 
\end{eqnarray}
are the contributions from 
one-loop subdivergences
and from the remaining two-loop divergences, respectively.
\end{widetext}
Also,
\begin{eqnarray}
T(x,y,z) &=& -\frac{\partial}{\partial x} S(x,y,z) .
\end{eqnarray}
Similarly, define
\begin{eqnarray}
&&U(x,y,z,u) = \nonumber \\ 
&&\lim_{\epsilon \rightarrow 0} \left [
{\bf U}(x,y,z,u)- U^{(1)}_{\rm div} (x,y) - U^{(2)}_{\rm div} 
\right ]
\end{eqnarray}
where
\begin{eqnarray}
U^{(1)}_{\rm div} (x,y) &=& {\bf B}(x,y)/\epsilon ,
\\
U^{(2)}_{\rm div} &=& -1/2\epsilon^2 + 1/2\epsilon
\end{eqnarray}
and, since the master integral is free of divergences,
\begin{eqnarray}
M(x,y,z,u,v) &=&
\lim_{\epsilon \rightarrow 0} {\bf M}(x,y,z,u,v) .
\end{eqnarray}
Thus, the bold-faced letters $\bf A,B,S,T,U$ represent the original 
regularized integrals that diverge as $\epsilon \rightarrow 0$, while the 
ordinary letters $A,B,S,T,U,M$ are finite and independent of $\epsilon$
by definition.
Also, note that these integrals have various symmetries that are clear
from the diagrams: 
\begin{itemize}
\item $S(x,y,z)$ is invariant under interchange of any two of $x,y,z$.
\item $T(x,y,z)$ is invariant under $y \leftrightarrow z$.
\item $U(x,y,z,u)$ is invariant under $z \leftrightarrow u$.
\item $M(x,y,z,u,v)$ is invariant under the interchanges
$(x,z)\leftrightarrow (y,u)$, 
and 
$(x,y)\leftrightarrow (z,u)$, 
and 
$(x,y)\leftrightarrow (u,z)$. 
\end{itemize}
This leads to many obvious permutations on formulas given below, which
will not be noted explicitly.

It is useful to define several related functions. The two-loop vacuum
integral is
\begin{eqnarray}
I(x,y,z) &=& S(x,y,z) |_{s = 0} .
\end{eqnarray}
It is equal to $(16 \pi^2)^2$ times the integral
$\hat I(x,y,z)$ in \cite{Ford:pn}
and is precisely equal to the same function used in 
\cite{Martin:2001vx}. In the 
present paper, the analytical expression 
is reviewed in section 
\ref{analytical} and the recurrence relation for
derivatives in section \ref{exps}.

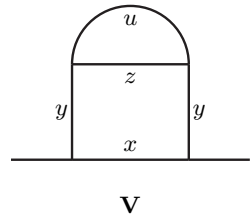
\begin{figure}
\begin{picture}(108,80)(-54,-40)
\SetWidth{0.9}
\Line(-45,-18)(-22,-18)
\Line(45,-18)(22,-18)
\Line(-22,-18)(22,-18)
\CArc(0,18)(22,0,180)
\Line(-22,18)(22,18)
\Line(-22,-18)(-22,18)
\Line(22,-18)(22,18)
\Text(0,-13)[]{$x$}
\Text(-26,0)[]{$y$}
\Text(26,0)[]{$y$}
\Text(0,13)[]{$z$}
\Text(0,35)[]{$u$}
\Text(0,-35)[]{$\bf V$}
\end{picture}
\caption{\label{fig:V} The two-loop Feynman diagram for $V(x,y,z,u)$.
%which is equal to $-\partial U(x,y,z,u)/\partial y$.
}
\end{figure}

The integral $T(x,y,z)$ has a 
logarithmic
infrared divergence as $x\rightarrow 0$. This divergence must cancel from 
physical
quantities, but as a book-keeping device it is useful to have a 
version of the integral $T(0,x,y)$ with the infrared divergence removed:
\begin{eqnarray}
\Tbar (0,x,y) &=& \lim_{\delta \rightarrow 0} \left [
T(\delta,x,y) + B(x,y) \lnbar\delta 
\right ].
\end{eqnarray}
Finally, for future reference we note that the topology $V$ in
Figure \ref{fig:V} arises quite often. 
When the vertical propagators are different, the result of the diagram
is just the difference of two $U$ functions. However, when the 
vertical
propagators have 
the
same squared mass $y$, it is useful to define the corresponding
integral
\begin{eqnarray}
V(x,y,z,u) = -\frac{\partial}{\partial y} U(x,y,z,u) .
\label{defV}
\end{eqnarray}
In section \ref{mderivs}, I will provide the formula expressing
$V(x,y,z,u)$ algebraically in terms of the other basis integrals.

To illustrate the usefulness of the above definitions, consider
the most general renormalizable theory of real scalar
fields $\phi_i$, governed by the interaction Lagrangian
\begin{eqnarray}
{\cal L} = 
-\frac{1}{2} m^2_i \phi_i^2 
- \frac{\lambda^{ijk}}{6} \phi_i \phi_j \phi_k
- \frac{\lambda^{ijkn}}{24} \phi_i \phi_j \phi_k \phi_n . \phantom{xxx}
\end{eqnarray}
Here $m^2_i$, $\lambda^{ijk}$ and $\lambda^{ijkn}$ are the tree-level 
renormalized masses and couplings.
Then, defining the self-energy matrix function $\Pi_{ij}(s)$ so that
the pole masses and widths $M,\Gamma$
are the solutions\footnote{This equation should be solved in
the Taylor series sense; the self-energy and its derivatives are
first evaluated only for $s$ with an infinitesimal positive imaginary 
part. That data is then used to construct a Taylor series expansion 
for complex $s$. This is necessary because 
the imaginary part of the pole mass is negative, while
the standard convention (as here) is that the infinitesimal imaginary 
part of the physical-sheet $s$ is positive.} 
for complex $s = M^2 - i \Gamma M$ of the eigenvalue equation
\begin{eqnarray}
(s - m^2_i) \delta_{ij} - \Pi_{ij}(s) = 0,
\end{eqnarray}
one has:
\begin{eqnarray}
\Pi_{ij}(s) = \frac{1}{16 \pi^2} \Pi^{(1)}_{ij}(s) + 
\frac{1}{(16 \pi^2)^2} \Pi^{(2)}_{ij}(s) + \ldots,
\end{eqnarray}
with 
\begin{widetext}
\begin{eqnarray}
\Pi^{(1)}_{ij}(s) &=& \frac{1}{2} \lambda^{ijkk} A(m_k^2) 
-\frac{1}{2} \lambda^{ikn} \lambda^{jkn} B(m_k^2,m_n^2) , \\
\Pi^{(2)}_{ij}(s) &=& 
-\frac{1}{2} \lambda^{ikn} \lambda^{jmp} \lambda^{kmr} \lambda^{npr}
M(m_k^2, m_m^2, m_n^2, m_p^2, m_r^2) 
\nonumber
\\ && 
+ \frac{1}{2} \lambda^{ikm} \lambda^{jkn} \lambda^{mpr} \lambda^{npr}
[U(m_k^2, m_m^2, m_p^2, m_r^2) 
-U(m_k^2, m_n^2, m_p^2, m_r^2)]/[m_m^2 - m_n^2] 
\nonumber
\\ && 
+ 
\frac{1}{2} \left [ \lambda^{ikm} \lambda^{jknp} \lambda^{mpn}
U(m_k^2, m_m^2, m_n^2, m_p^2) + (i \leftrightarrow j) \right ]
- 
\frac{1}{6} \lambda^{ikmn} \lambda^{jkmn} S(m_k^2,m_m^2,m_n^2)
\nonumber
\\ && 
+ 
\frac{1}{4} \lambda^{ikm} \lambda^{jnp} \lambda^{kmnp} 
B(m_k^2,m_m^2) B(m_n^2,m_p^2)
+ 
\frac{1}{4} \lambda^{ijkm} \lambda^{kmnn} 
A(m_n^2) [A(m_k^2) - A(m_n^2)]/[m_k^2 - m_n^2]
\nonumber
\\ && 
+ 
\frac{1}{2} \lambda^{ikm} \lambda^{jkn} \lambda^{mnpp}
A(m_p^2) [B(m_k^2,m_m^2) - B(m_k^2,m_n^2)]/[m_n^2 - m_m^2]
\nonumber
\\ && 
+ 
\frac{1}{4} \lambda^{ijkm} \lambda^{knp} \lambda^{mnp}
[I(m_k^2,m_n^2,m_p^2) - I(m_m^2,m_n^2,m_p^2)]/[m_m^2 - m_k^2],
\end{eqnarray}
in which the \MSbar~counterterms have been included.
(Note that for degenerate masses, the function $V$ will appear,
as well as derivatives of the functions $A,B,I$.) 
Of course, for theories involving fermions and vectors, things
are more complicated, but the basis functions as defined above
tend to neatly organize the counterterms, at least in mass-independent
renormalization schemes.

In the following, a prime on a squared-mass argument of a function
stands for a
derivative with respect to that argument. This notation is 
particularly
convenient when there are many derivatives or when some of the
arguments are set equal after differentiation.
Thus, for example,
\begin{equation}
f(x'',x,y') \equiv \lim_{z\rightarrow x}
\left [\frac{\partial^3}{\partial x^2 \partial y} f(x,z,y) \right ] .
\end{equation}

Several kinematic shorthand notations used throughout this paper are:
\begin{eqnarray}
\Delta_{xyz} &=& x^2 + y^2 + z^2 - 2 x y -2 x z -2 y z ,
\\
D_{sxyz} &=& 
s^4 - 4 s^3 (x+y+z) 
+ s^2 [4(x + y + z)^2 + 2 \Delta_{xyz}]
-s [64 x y z + 4 (x+y+z)\Delta_{xyz}]
+ \Delta_{xyz}^2 ,
\\
\Delta &=& 
s^2 v + s [v (v-u-x-y-z) + (x-y)(z-u)] + (u x - y z)(u+x-y-z)+ v 
(x-z)(y-u) .
\end{eqnarray}

\section{Derivatives of basis integrals with respect to squared-mass 
arguments\label{mderivs}}
\setcounter{equation}{0}

In this section, I present the results of taking derivatives of the basis
integrals with respect to squared-mass arguments. These can be obtained
straightforwardly, if tediously, from Tarasov's algorithm. The necessary
recurrence relations have been implemented by Mertig and Scharf in the
computer algebra program TARCER \cite{Mertig:1998vk}, which was used
to derive or check most of the results in this section. The
results below for (the equivalents of) the $S$ and $T$ functions have 
already
been given in \cite{Caffo:1998du}.

For the one-loop self-energy integral, one has:
\begin{eqnarray}
\frac{\partial}{\partial x} B(x,y) &=&
\frac{1}{\Delta_{sxy}} \left [
(x-y-s) (B(x,y)-2) + (x+y-s) \lnbarx -2 y \lnbary
\right ].
\label{dBdx}
\end{eqnarray}
Derivatives of the sunrise function $S$ are trivial, in the sense that 
they are 
already
included in the basis:
\begin{eqnarray}
\frac{\partial}{\partial x} S(x,y,z) &=& -T(x,y,z) .
\label{dSdx}
\end{eqnarray}
For the $T$ function, there are two distinct derivatives. First,
\begin{eqnarray}
\frac{\partial}{\partial x} T(x,y,z) &=& \frac{1}{x D_{sxyz}} \left [
k_{TxS} S(x,y,z) + k_{TxT1} T(x,y,z) + k_{TxT2} T(y,x,z)
+ k_{TxT3} T(z,x,y) 
+ k_{Tx} \right ],
\label{dTdx}
\end{eqnarray}
where the coefficient functions are 
\begin{eqnarray}
k_{TxS} &=& 
-2s^3 + 6 s^2 (x+y+z) + s [2\Delta_{xyz}- 8 (x^2+y^2+z^2)]
+ 2(x+y+z)\Delta_{xyz} + 32 xyz 
\\
k_{TxT1} &=& 2x(x-s)\left [
s^2 - 2 s (x+y+z) + \Delta_{xyz} + 8 y z \right ]
\\
k_{Tx} &=& \Bigl \lbrace 5 s^4/12 + s^3 x[\lnbarx- 27/4 ]
+ s^2 x [ \lnbarx ( y \lnbary + z \lnbarz -3x - 7y- 7z)+51 x/4 + 53(y+z)/4  ]
\nonumber \\ &&
+ s x[  \lnbarx \lbrace 2 (z-x-y) y \lnbary +2 (y-x-z) z \lnbarz
+3 x^2 +10 x (y+z) +11 (y^2 + z^2) - 14 y z \rbrace-41 x^2/4 
\nonumber \\ &&
-103 x (y+z)/4 + 11 y z/6 ]
+ x \lnbarx \lbrace y \lnbary [ (x-y)^2 + 2 z (x+y) -3 z^2]
+  z \lnbarz [(x-z)^2 + 2 y (x+z) -3 y^2]
\nonumber \\ &&
+ x (9 y^2 + 9z^2 - 26 y z - x^2) - 3 x^2 (y+z) 
-5 (y+z) (y-z)^2 \rbrace + 3 x^2 [(x+y+z)^2 -4 (y-z)^2]
\Bigr \rbrace
\nonumber \\ && 
+ \Bigl \lbrace (x \leftrightarrow y)\Bigr \rbrace
+ \Bigl \lbrace (x \leftrightarrow z)\Bigr \rbrace ,
\end{eqnarray}
and $k_{TxT2}$ is obtained from $k_{TxT1}$ by $(x \leftrightarrow y)$,
and $k_{TxT3}$ is obtained from $k_{TxT1}$ by $(x \leftrightarrow z)$.
The symmetries of the preceding expressions imply that
\begin{eqnarray}
x \frac{\partial^2}{\partial x^2} S(x,y,z) =
y \frac{\partial^2}{\partial y^2} S(x,y,z) ,
\end{eqnarray}
an identity which seems somewhat remarkable since it is not immediately 
obvious from the symmetries of the Feynman diagram.
When $z=0$, this simplifies to:
\begin{eqnarray}
x S(x'',y,0) = y S(x,y'',0) = B(x,y).
\end{eqnarray}
The other derivative of the $T$ function is given by
\begin{eqnarray}
\frac{\partial}{\partial y} T(x,y,z) &=& \frac{1}{D_{sxyz}}
\left [ k_{TyS} S(x,y,z)
+ k_{TyT1} T(x,y,z) 
+ k_{TyT2} T(y,x,z)
+ k_{TyT3} T(z,x,y) 
+ k_{Ty}  
\right ]  ,
\label{dTdy}
\end{eqnarray}
where
\begin{eqnarray}
k_{TyS} &=& 
-4s^2 + 8 s (x+y-z) + 12 z^2 - 8 z (x+y) - 4(x-y)^2 
\\
k_{TyT1} &=& 
-s^3 + s^2 (z+3y-x) 
+s (5 x^2 + 6 x y-3 y^2-14 x z + 2 y z + z^2)
-3 x^2 (x+z)+ (y-z)^3 
+ 7 x^2 y
\nonumber \\ &&
+ 7 x z^2 - 5 x y^2 - 2 x y z 
\\
k_{TyT3} &=& -8 s^2 z +8 s z (x+y) +8z^2 (z-x-y)
\\
k_{Ty} &=& \Bigl \lbrace
s^3 [\frac{1}{2}\lnbarx \lnbary - 2 \lnbarx +\frac{11}{4}]
+ s^2 [ 
(2 \lnbarx -3) z \lnbarz 
-3 (x+z/2) \lnbarx \lnbary
+ (8x+6y+2z) \lnbarx
-20x +z ]
\nonumber \\ &&
+ s [
(3 x^2 + x y + 2 x z + 3 z^2/2) \lnbarx\lnbary
+4 z (x-y-z) \lnbarx \lnbarz + z(10z-4x)\lnbarz 
+ 2 (z^2 - 3 y^2 - 5 x^2 - 4 x y 
\nonumber \\ &&
- 4 x z + 2 y z) \lnbarx
+ 47 x^2/2 + 25 x y/2 + 11 x z - 69 z^2/4 ]
+ ( x^2 (y+z) - x^3 - 5 x y z + x z^2 - z^3/2) \lnbarx \lnbary
\nonumber \\ &&
+ 2 z (y^2 + z^2 - 3 x^2 + 2 x (y+z) - 2 y z) \lnbarx \lnbarz
+ z (10 x^2 - 10 x y + 4 x z - 7 z^2) \lnbarz 
+ 2 (2 x^3 - 3 x^2 y + y^3 
\nonumber \\ &&
+ 7 x^2 z + 8 x y z -3 y^2 z - 8 x z^2 + 3 y z^2 - z^3) \lnbarx
+ 9 (x z^2 - x^3 + x^2 y - 3 x^2 z - x y z + 3 z^3/2)
\Bigr \rbrace
\nonumber \\ && 
+ (x\leftrightarrow y) ,
\end{eqnarray}
and $k_{TyT2}$ is obtained from $k_{TyT1}$ by $(x\leftrightarrow y)$.
For the special case of a vanishing first argument, one finds
\begin{eqnarray}
\frac{\partial}{\partial x} \Tbar (0,x,y) &=&
k_{\Tbar S} S(0,x,y)
+ k_{\Tbar \Tbar} \Tbar (0,x,y)
+ k_{\Tbar T1} T (x,0,y)
+ k_{\Tbar T2} T (y,0,x)
+ k_{\Tbar}
\label{dTbardx}
\end{eqnarray}
where
\begin{eqnarray}
k_{\Tbar S} &=&
16 y (y-x-s)/\Delta_{sxy}^2 -4/\Delta_{sxy}
\\
k_{\Tbar \Tbar} &=&
(x-y-s)/\Delta_{sxy}
\\
k_{\Tbar T1} &=&
8 x y (y-x-3s)/\Delta_{sxy}^2 -(s+3x+y)/\Delta_{sxy}
\\
k_{\Tbar T2} &=&
8 y [-s^2 + s x -y x + y^2]/\Delta_{sxy}^2
\\
k_{\Tbar } &=& 
\frac{2 y}{\Delta_{sxy}^2} \Bigl [ s [4 x \lnbarx (\lnbary -1) 
+(4y-8x) \lnbary + 15 x-7y]
+ (x-y) [ 4 x \lnbarx (3-\lnbary) 
+ (8x + 4y) \lnbary
\nonumber \\ &&
-29x -7 y ] \Bigr ]
+ \left [
s (11/2 - 2 \lnbarx) +
2 y \lnbary (\lnbarx-3) + (4x-2y) \lnbarx - 9x +13 y
\right ]/\Delta_{sxy} .
\end{eqnarray}

The derivatives of the $U$ functions are:
\begin{eqnarray}
\frac{\partial}{\partial x} U(x,y,z,u) &=&
\frac{1}{\Delta_{sxy}} \Bigl [
(x-y-s) U(x,y,z,u) + 2 z T(z,u,x) + 2 u T(u,x,z)
+ (3 x -y+s) T(x,z,u) 
\nonumber \\ &&
+ 4 S(x,z,u) -2 I(y,z,u)
-2 (A(u)+A(x)+A(z)) + 2 (x+z+u) 
-s/2 \Bigr ]
\label{dUdx}
\\
\frac{\partial}{\partial z} U(x,y,z,u) &=&
\frac{1}{\Delta_{yzu}} \Bigl [
(z-y-u) U(x,y,z,u) + (u+z-y) T(z,x,u) - 2 u T(u,x,z)
+u+y-z
\nonumber \\ &&
+ [(u+z-y)\lnbarz + 2 u(1-\lnbaru) +2y-2z ] B(x,y)
\Bigr ]
\label{dUdz}
\\
\frac{\partial}{\partial y} U(x,y,z,u) &=&
k_{UU} U(x,y,z,u) + k_{UT1} T(x,z,u) + k_{UT2} T(u,x,z) 
+ k_{UT3} T(z,x,u)
\nonumber \\ && 
+ k_{US} [ S(x,z,u) -(A(x)+A(z)+A(u)+ I(y,z,u))/2]
+ k_{UB} B(x,y)
+ k_U 
\label{dUdy}
\end{eqnarray}
where the coefficient functions in the last expression are
\begin{eqnarray}
k_{UU} &=& (y-x-s)/\Delta_{sxy} + (y-z-u)/\Delta_{yzu} -1/y 
\\
k_{UT1} &=& 2 x (s-x)/y\Delta_{sxy}
\\
k_{UT2} &=& 
u (s-x-y)/y\Delta_{sxy}
+ u (y+z-u)/y\Delta_{yzu} 
\\
k_{US} &=& 2 (s-x-y)/y\Delta_{sxy}
\\
k_{UB} &=& [(y+z-u) u\lnbaru + (y+u-z) z \lnbarz 
            + (u-z)^2  -y^2]/y\Delta_{yzu}
\\
k_{U} &=& 
[-s^2/4 + s (z + u+ 5 x/4 + y/4)- (z+u+x)(x+y)]/y\Delta_{sxy}
+ (u+z-y)/\Delta_{yzu} 
\end{eqnarray}
and $k_{UT3}$ is related to $k_{UT2}$ by $(z \leftrightarrow u)$.
Some care is needed in treating cases where the 
denominator $\Delta_{yzu}$ threatens to vanish. One finds by taking the
limits that
\begin{eqnarray}
U(x,0,y,y') &=& T(y,y',x)/2 - T(y',y,x)/2 -B(0,x)/2y
\\
U(x,y,y',0) &=& \left [
\Tbar (0,x,y) - T(y,0,x) - B(x,y) \lnbary \right ]/2y
\\
U(x,y',y,0) &=& -U(x,y,y',0) + (2-\lnbary ) B(x,y') .
\label{dUdyy0}
\end{eqnarray}

There are two types of derivatives of the master integral function $M$.
First, 
\begin{eqnarray}
\frac{\partial}{\partial x} M(x,y,z,u,v) &=& 
k_{MxU1} U(x,z,u,v) +k_{MxU2} U(y,u,z,v)
+k_{MxU3} U(z,x,y,v) +k_{MxU4} U(u,y,x,v)
\nonumber \\ &&
+ k_{MxS} \left [ S(x,u,v)+ S(y,z,v) +\frac{s}{2}B(x,z)B(y,u)
-\frac{1}{2} I(x,y,v) -\frac{1}{2} I(z,u,v) \right ]
\nonumber \\ &&
+ k_{MxT1} T(x,u,v) + k_{MxT2} T(y,z,v)
+ k_{MxT3} T(z,y,v) + k_{MxT4} T(u,x,v)
\nonumber \\ &&
+ k_{MxT5} [T(v,x,u) + T(v,y,z)]+ k_{MxB1} B(x,z) + k_{MxB2} B(y,u) + k_{Mx}
\label{dMdx}
\end{eqnarray}
where the coefficient functions are
\begin{eqnarray}
k_{MxU1} &=& \frac{z}{\Delta_{sxz} \Delta} \Bigl [
           s^2 + s (2v-x-y-z-u) + (x-z)(y-u) \Bigr ]
\\
k_{MxU2} &=& -u/\Delta
\\
k_{MxU3} &=& \frac{v-u}{\Delta} 
+ \frac{1}{\Delta_{sxz} \Delta} \Bigl [s (v x + v z + 2 x z - y z - u x) +
(x - z) (u x - v x + v z - y z) \Bigr ]
\nonumber \\ &&
+ \frac{1}{\Delta_{xyv} \Delta} \Bigl [
s v (v-x-y) -u v x   + u x^2 - u x y + 2 v x y - v y z -x y z + y^2 z 
\Bigr ]
\\
k_{MxU4} &=& \frac{y}{\Delta_{xyv}\Delta} \Bigl [
2 s v- u v + v^2 - u x - v x + u y - v y - v z + x z - y z \Bigr ]
\\
k_{MxS} &=& \frac{2}{\Delta_{sxz}\Delta} \Bigl [
s (u-z-v) -u x + v x - u z - v z - x z + 2 y z + z^2 \Bigr ]
\\
k_{MxT1} &=& x k_{MxS}/2 - k_{MxU3}
\\
k_{MxT2} &=& y k_{MxS}/2 - k_{MxU4}
\\
k_{MxT3} &=& z k_{MxS}/2 - k_{MxU1}
\\
k_{MxT4} &=& u k_{MxS}/2 - k_{MxU2}
\\
k_{MxT5} &=& v k_{MxS}/2 + \frac{v}{\Delta_{xyv}\Delta}\Bigl [
s (v-x+y) + u (x+y-v) + y (x-y-2z+v) \Bigr ]
\\
k_{MxB1} &=& \frac{u}{\Delta} \lnbaru - k_{MxU4} \lnbary +
(k_{MxT5}-v k_{MxS}/2) \lnbarv + \frac{2}{\Delta_{xyv}\Delta} \Bigr [
s v (x+y-v) + u v x - u x^2  + u x y 
\nonumber \\ &&
- 2 v x y 
+ v y z + x y z - y^2  z  \Bigl ]
\\
k_{MxB2} &=& \frac{2}{\Delta} (v+z) - k_{MxU3} \lnbarx- k_{MxU1} \lnbarz
+ (k_{MxT5} - v k_{MxS}/2) \lnbarv - 2 k_{MxU4} 
\nonumber \\ &&
+ \frac{2}{\Delta_{sxz}\Delta} \Bigl [
s (v x - u x - u z + 3 v z + 3 x z - 2 y z+z^2 ) + (x-z)^2 (u-v-z) \Bigr ]
\\
k_{Mx} &=& -k_{MxS}
( x \lnbarx + y \lnbary + z \lnbarz + u \lnbaru +2 v \lnbarv )/2
+ \frac{1}{2\Delta}(u+z+v-2y) + \frac{1}{\Delta_{xyv}\Delta} [s v(x-y-v) 
\nonumber \\ &&
+v(u x + u y - 3 x y - y^2  + 2 y z) +(y - u) (x - y)^2 ]
+ \frac{1}{2 \Delta_{sxz}\Delta} \Bigl [
s[ 4 u (u+ v+y) - 8 v^2 + 3 u x - 3 v x  
\nonumber \\ &&
- 4 v y - u z - 9 v z - x z - 6 y z - 3 z^2]
+ 8 v^2 (x-z) + x^2 (3 v -3 u - 5 z) + z^2 (u + 3 v + 2 x + 12 y)
\nonumber \\ &&
- 4 u^2 (x+z) + 4 (v+z-u) x y +6 (2 y - 2 u -x) v z
-4 u v x - 14 u x z + 4 u y z  + 8 z y^2 + 3 z^3 
\Bigr ] .
\end{eqnarray}
Finally, 
\begin{eqnarray}
\frac{\partial}{\partial v} M(x,y,z,u,v) &=&
k_{MvU1} U(x,z,u,v) + k_{MvU2} U(y,u,z,v)
+k_{MvU3} U(z,x,y,v) +k_{MvU4} U(u,y,x,v)
\nonumber \\ &&
+ k_{MvS} \left [ S(x,u,v)+ S(y,z,v) +\frac{s}{2}B(x,z)B(y,u)
-\frac{1}{2} I(x,y,v) -\frac{1}{2} I(z,u,v) \right ] 
\nonumber \\ &&
+ k_{MvT1} T(x,u,v) + k_{MvT2} T(y,z,v) 
+ k_{MvT3} T(z,y,v) + k_{MvT4} T(u,x,v)
\nonumber \\ &&
+ k_{MvT5} [T(v,x,u) + T(v,y,z)]
+ k_{MvB1} B(x,z) + k_{MvB2} B(y,u) + k_{Mv}
\label{dMdv}
\end{eqnarray}
where 
\begin{eqnarray}
k_{MvU1} &=& \frac{z}{\Delta_{uzv}\Delta} \Bigl [
s (z-u-v) + u^2 - u v + 2 u x - u y + v y - u z - y z \Bigr ]
\\
k_{MvS} &=& -2/\Delta
\\
k_{MvT1} &=& -k_{MvU3} - x/\Delta
\\
k_{MvT5} &=& -(s+v)/\Delta + k_{MvU1} + k_{MvU2} + k_{MvU3} + k_{MvU4}
\\
k_{MvB1} &=& (k_{MvT5} + v/\Delta) \lnbarv - k_{MvU4} \lnbary 
- k_{MvU2} \lnbaru + 2 s/\Delta - 2 k_{MvU1} - 2 k_{MvU3}
\\
k_{Mv} &=& \frac{1}{\Delta} \left [
x \lnbarx + y \lnbary + z \lnbarz +u \lnbaru +
2 v \lnbarv -2 (x+y+z+u)-5v+s/2 \right ] -k_{MvT5} .
\end{eqnarray}
Here
$k_{MvU2}$, $k_{MvT2}$, $k_{MvB2}$ 
are each respectively related to 
$k_{MvU1}$, $k_{MvT1}$, $k_{MvB1}$ by 
$(x,z) \leftrightarrow (y,u)$. 
Similarly, 
$k_{MvU3}$, $k_{MvT3}$
are each related to $k_{MvU1}$, $k_{MvT1}$ by
$(x,y)\leftrightarrow (z,u)$,
and 
$k_{MvU4}$, $k_{MvT4}$ are related to $k_{MvU1}$, $k_{MvT1}$ by
$(x,y)\leftrightarrow (u,z)$. 

By repeatedly applying the identities in this section, one may obtain
the results for two-loop Feynman self-energy integrals with arbitrary
powers of propagators in the denominator. An important example is that
equations~(\ref{dUdy}) and (\ref{dUdyy0}) can be used to
find the integral $V(x,y,z,u)$ defined in eq.~(\ref{defV}) and 
corresponding
to the topology shown in Figure \ref{fig:V}.

\end{widetext}

\section{Differential equations in the external momentum invariant 
$s$\label{diffeqs}}
\setcounter{equation}{0}

In this section, I present results for the derivatives of the
basis functions with respect to $s$. These are most easily
obtained by dimensional analysis, using the facts that $B$, $S$, 
$T$, $\Tbar$, $U$, and $M$ have mass dimensions 0, 2, 0, 0, 0, and $-2$
respectively. Since the only dimensionful quantities on which they
depend are $Q^2$, $s$, and the propagator masses, we have:
\begin{eqnarray}
\sum_{\alpha = Q^2,s,x,y} \alpha \frac{\partial}{\partial \alpha}
B(x,y) &=& 0 ,
\label{dBdalpha}
\\
\sum_{\alpha = Q^2,s,x,y,z} \alpha \frac{\partial}{\partial \alpha} 
S(x,y,z) &=& S(x,y,z) ,
\label{dSdalpha}
\\
\sum_{\alpha} \alpha \frac{\partial}{\partial \alpha}
T(x,y,z) &=& 0 ,
\label{dTdalpha}
\\
\sum_{\alpha} \alpha \frac{\partial}{\partial \alpha}
\Tbar (0,x,y) &=& 0 ,
\label{dTbardalpha}
\\
\sum_{\alpha} \alpha \frac{\partial}{\partial \alpha}
U(x,y,z,u) &=& 0 ,
\label{dUdalpha}
\\
\sum_{\alpha} \alpha \frac{\partial}{\partial \alpha}
[s M(x,y,z,u,v)] &=& 0 ,
\label{dMdalpha}
\end{eqnarray}
where in each case $\alpha$ is summed over $Q^2$, $s$, and the appropriate
$x,y,\ldots$. Section \ref{mderivs} already gave the derivatives with 
respect to
the squared masses. The derivatives with respect to the renormalization 
scale are easily obtained from the definitions in section 
\ref{conventions}:
\begin{eqnarray}
Q^2 \frac{\partial}{\partial Q^2} A(x) &=& -x
,
\\
Q^2 \frac{\partial}{\partial Q^2} B(x,y) &=& 1
,
\label{dBdQ}
\\
Q^2 \frac{\partial}{\partial Q^2} S(x,y,z) &=& 
A(x)+A(y)+A(z)
\nonumber \\ &&
-x-y-z +s/2
,
\label{dSdQ}
\\
Q^2 \frac{\partial}{\partial Q^2} T(x,y,z) &=& -A(x)/x
,
\label{dTdQ}
\\
Q^2 \frac{\partial}{\partial Q^2} \Tbar (0,x,y) &=& 1 - B(x,y)
,
\label{dTbardQ}\
\\
Q^2 \frac{\partial}{\partial Q^2} U(x,y,z,u) &=&
1+B(x,y)
, 
\label{dUdQ}
\\
Q^2 \frac{\partial}{\partial Q^2} M(x,y,z,u,v) &=& 0 .
\label{dMdQ}
\end{eqnarray}
Now, combining equations (\ref{dBdx}), (\ref{dBdalpha}), and (\ref{dBdQ}), 
one finds
\begin{widetext}
\begin{eqnarray}
s \frac{d}{ds} B(x,y) &=& \frac{1}{\Delta_{sxy}} \Bigl [
(s (x+y)-(x-y)^2) B(x,y) + (s-x+y) A(x) + (s+x-y) A(y) +s (x+y-s)
\Bigr ].
\label{dBds}
\end{eqnarray}
Similarly, combining equations (\ref{dSdx}), (\ref{dSdalpha}), and 
(\ref{dSdQ}),
one gets the result for the sunrise function
\begin{eqnarray}
s \frac{d}{ds} S(x,y,z) &=&
S(x,y,z) + x T(x,y,z) + y T(y,x,z) + z T(z,x,y) 
-A(x)-A(y)-A(z) 
\nonumber \\ &&
+x+y+z -s/2 ,
\label{dSds}
\end{eqnarray}
and, from (\ref{dTdx}), (\ref{dTdy}), (\ref{dTdalpha}), and (\ref{dTdQ}):
\begin{eqnarray}
s \frac{d}{ds} T(x,y,z) &=& 
c_{TS} S(x,y,z)
+ c_{TT1} T(x,y,z)
+ c_{TT2} T(y,x,z)
+ c_{TT3} T(z,x,y)
+ c_{T}
\label{dTds}
\end{eqnarray}
where
\begin{eqnarray}
c_{TS} &=& 
\frac{2}{D_{sxyz}} \left [
s^3 
- s^2 (3x+y+z) 
+ s (3 x^2 -y^2 - z^2 - 2 x y -2 x z + 10 y z) 
+ (y+z-x) \Delta_{xyz} \right ]
\\ 
c_{TT1} &=& 
\frac{1}{D_{sxyz}} \Bigl [
s^3 [2 x + y + z] -
s^2 [6x^2 + 3y^2 + 3z^2+ 3xy + 3xz + 2yz ]
+ s [  6x^3 + 3y^3 + 3z^3
\nonumber
\\ &&
- 5x^2 (y+z) 
- 4 x (y^2 + z^2) -3y z(y+z) + 40 x y z ]
+x (y+z-x) \Delta_{xyz}- \Delta_{xyz}^2
\Bigr ]
\\ 
c_{TT2} &=& 
\frac{y}{D_{sxyz}} \left [
3 s^3 + s^2 (3z-7x-5y) + s (
5x^2 + y^2 - 7z^2- 6xy + 2xz + 14yz )
+ (y+z-x) \Delta_{xyz}
\right ]
\\
c_{T} &=& \frac{1}{D_{sxyz}} \Bigl \lbrace 
s^4 [(\lnbarx)/2 - 9/8] 
+ s^3 [
-y \lnbarx \lnbary -(2y+5x/2 )\lnbarx +y \lnbary + 43 x/8 + 21 y/4] 
\nonumber 
\\ &&
+s^2 [
y(x+3y +z) \lnbarx \lnbary -3 y z \lnbary \lnbarz
+ (3 x y + 9 x^2/2) \lnbarx +
y (x-5y+11z) \lnbary
-75 x^2/8 -21 x y/2 
\nonumber 
\\ &&
+ 5 y^2/4 -69 y z/4]
+s [
y (x^2 + 2 x y - 3 y^2 - 10 x z + 2 y z + z^2) \lnbarx \lnbary
+ 2 y z (x + 2y) \lnbary \lnbarz
\nonumber 
\\ &&
+ (4 x^2 y + x y^2 + 2 y^3 - 2 y^2 z - 5 xyz -7 x^3/2) \lnbarx
+ y (7 y^2 - 5 x^2 - 2 x y  + 14 x z - 22 y z- 9 z^2) \lnbary
\nonumber 
\\ &&
+ (57 x^3/2 - 7 x^2 y - 13 x y^2 - 37 y^3 + 25 x y z + 181 y^2 z)/4 
]
+ \Delta_{xyz} [ 
y (y-x-z) \lnbarx \lnbary 
\nonumber 
\\ &&
+ y z \lnbary \lnbarz
+ (x^2 -x y - y^2 + y z) \lnbarx +
3y (x-y-z)\lnbary - 2 x^2 - x y + 5 y^2 +4 y z]
\Bigr \rbrace + (y \leftrightarrow z)
\end{eqnarray}
and $c_{TT3}$ is obtained from $c_{TT2}$ 
by the interchange $(y \leftrightarrow z)$.
The equivalents of equations (\ref{dSds}) and (\ref{dTds}) 
were found earlier in \cite{Caffo:1998du}.

For the $\Tbar$ function, I find from equations (\ref{dTbardx}), 
(\ref{dTbardalpha}), and (\ref{dTbardQ}),
\begin{eqnarray}
s \frac{d}{ds} \Tbar(0,x,y) &=&
c_{\Tbar \Tbar} \Tbar(0,x,y) +c_{\Tbar T1} T(x,0,y)
+c_{\Tbar T2} T(y,0,x) +c_{\Tbar S} S(0,x,y) +c_{\Tbar}
\label{dTbards}
\end{eqnarray}
where
\begin{eqnarray}
c_{\Tbar \Tbar} &=& [s (x+y) - (x-y)^2 ]/\Delta_{sxy}
\\
c_{\Tbar T1} &=& x (3s+x+9y)/\Delta_{sxy}
+ 8 x y [s (5x+y)-(x-y)^2 ]/\Delta_{sxy}^2
\\
c_{\Tbar S} &=& 2 (s+x+y)/\Delta_{sxy} + 32 s x y/\Delta_{sxy}^2
\\
c_{\Tbar} &=& \frac{1}{\Delta_{sxy}^2} \Bigl [
-9 s^4/8 + s^3 x (\lnbarx + 21/4) 
+ s^2 x [-3 y \lnbarx\lnbary + (11y-5x) \lnbarx + 5 x/4 - 69y/4]
+ s x [4 x y \lnbarx \lnbary 
\nonumber \\ &&
+ (7 x^2-22 x y-9 y^2) \lnbarx
- 37 x^2/4 + 181 x y/4 ]
+ (x-y)^2 x [ y \lnbarx \lnbary -3 (x+y) \lnbarx + 5 x + 4 y ]
\Bigr ] 
\nonumber \\ && + (x \leftrightarrow y),
\end{eqnarray}
and $c_{\Tbar T2}$ is obtained from $c_{\Tbar T1}$ by $x \leftrightarrow 
y$.

The differential equation for the $U$ function, obtained from
equations 
(\ref{dUdx}),
(\ref{dUdz}),
(\ref{dUdy}),
(\ref{dUdalpha}), and
(\ref{dUdQ}), is
\begin{eqnarray}
s \frac{d}{ds} U(x,y,z,u) &=& \frac{1}{\Delta_{sxy}} 
\Bigl ( [s (x+y) - (x-y)^2] U(x,y,z,u) 
+x (y-x-3s) T(x,u,z) + (y-x-s) [2 S(x,z,u) 
\nonumber
\\ &&
+ u T(u,x,z) + z T(z,x,u) -I(y,z,u) 
+x(2-\lnbarx)+z(2-\lnbarz)+u(2-\lnbaru) - s/4] \Bigr ) .
\label{dUds}
\end{eqnarray}
The equivalent of this result was obtained earlier in 
\cite{Caffo:1998yd}.

For the master integral $M$, I find from equations
(\ref{dMdx}),
(\ref{dMdv}),
(\ref{dMdalpha}), and
(\ref{dMdQ}) that:
\begin{eqnarray}
\frac{d}{ds} [s M(x,y,z,u,v)] &=& 
s \left [c_{MU1} U(x,z,u,v) 
+c_{MU2} U(y,u,z,v) 
+c_{MU3} U(z,x,y,v) 
+c_{MU4} U(u,y,x,v) \right ]
\nonumber
\\ &&
+ c_{MS} \Bigl [
S(x,u,v)+ S(y,z,v)
+ \frac{s}{2} B(x,z) B(y,u) -\frac{1}{2} I(x,y,v) -\frac{1}{2} I(z,u,v)
\Bigr ]
\nonumber
\\ &&
+ c_{MT1} T(x,u,v)
+ c_{MT2} T(y,z,v)
+ c_{MT3} T(z,y,v)
+ c_{MT4} T(u,x,v)
\nonumber
\\ &&
+ c_{MT5} [T(v,x,u) + T(v,y,z)]
+ c_{MB1} B(x,z)
+ c_{MB2} B(y,u)
+ c_{M}
\label{dMds}
\end{eqnarray}
where the coefficient functions are
\begin{eqnarray}
c_{MU1} &=& \frac{z}{\Delta_{sxz} \Delta} \left [
s (y-x-v)+ x^2 + 2ux - v x - x y + v z-x z - y z \right ]
\\
c_{MS} &=& 2 v/\Delta -2 (c_{MU1} +c_{MU2} +c_{MU3} +c_{MU4})
\\
c_{MT1} &=& 
x (v+z-u)/\Delta + x c_{MS}/2 -2x c_{MU1} - (x+z) c_{MU3}
\\
c_{MT5} &=& s v/\Delta + v c_{MS}/2 
\\
c_{MB1} &=& 
s v(\lnbarv -2)/\Delta + s c_{MU4} (2 - \lnbary) + s c_{MU2} (2 - \lnbaru)
\\
c_{M} &=& \left [(x-y)(u-z) -(3 s+x+y+z+u) v/2 \right ]/\Delta
+c_{MS} [2 v+x+y+z+u - v \lnbarv- (x \lnbarx +y \lnbary 
\nonumber \\ &&
+z \lnbarz +u \lnbaru)/2]
+[(3x+z) c_{MU1} + (3y+u) c_{MU2} +(3z+x) c_{MU3} +(3u+y) c_{MU4} ]/2 .
\end{eqnarray}
Here, the coefficient functions 
$c_{MU2}$, $c_{MT2}$, $c_{MB2}$ 
are each respectively related to 
$c_{MU1}$, $c_{MT1}$, $c_{MB1}$ by 
$(x,z) \leftrightarrow (y,u)$. 
Similarly, 
$c_{MU3}$, $c_{MT3}$
are each related to $c_{MU1}$, $c_{MT1}$ by
$(x,y)\leftrightarrow (z,u)$,
and 
$c_{MU4}$, $c_{MT4}$ are related to $c_{MU1}$, $c_{MT1}$ by
$(x,y)\leftrightarrow (u,z)$. 

\section{Expansions for small $s$\label{exps}}
\setcounter{equation}{0}

It is often useful to have expressions for the two-loop integral
functions expanded for small $s$. 
This provides the necessary initial data
for integrating the differential
equations numerically starting from $s=0$. 
The expansions, given in terms of the
analytically calculable vacuum function $I(x,y,z)$, can be obtained
by trying power series forms in the differential equations of
the previous section.

For example, for the one-loop function, one finds:
\begin{eqnarray}
B(x,y) &=&
\frac{A(y)-A(x)}{x-y} 
+ \frac{s}{2(x-y)^3} \Bigl [
x^2 - y^2 + 2 x y {\rm ln}(y/x) \Bigr ]
+ \frac{s^2}{6(x-y)^5} \Bigl [
(x-y)(x^2 + y^2 + 10 x y) 
\nonumber \\ &&
+ 6 x y (x+y) {\rm ln}(y/x)\Bigr ]
+ \ldots
,
\\
B(x,x) &=& -\lnbarx + \frac{s}{6x} + \frac{s^2}{60 x^2} + \ldots
.
\end{eqnarray}

For the two-loop functions, the most compact expressions involve
derivatives of the vacuum integral.
It is therefore useful to have a recurrence relation
for taking derivatives of the vacuum function $I(x,y,z)$:
\begin{eqnarray}
I(x',y,z) 
&=&
\frac{1}{\Delta_{xyz}} \Bigl [
(x-y-z) I(x,y,z) 
+ (x-y+z) A(x) A(y)/x
+ (x+y-z) A(x) A(z)/x
-2 A(y) A(z) 
\nonumber \\ && 
+(y+z-x) [A(x)+A(y)+A(z)]
+ x^2 -(y+z)^2 \Bigr ] ,
\\
I(x',x,0) &=& -\left (\lnbarx-1 \right )^2/2 ,
\\
I(x',0,0) &=& -\left (\lnbarx-1 \right )^2/2 - \zeta (2) .
\end{eqnarray}
These follow immediately from the analysis in \cite{Ford:pn}.
The function $I(x,y,z)$ obeys 
\begin{eqnarray}
x I(x',y,z) + y I(x,y',z) + z I(x,y,z') &=&
I(x,y,z) -A(x)-A(y)-A(z) +x+y+z ,
\\
x I(x'',y,z) &=& y I(x,y'',z) .
\end{eqnarray}
These identities make the presentation of the following formulas quite 
non-unique.

For the expansion of the sunrise integral, one finds
\begin{eqnarray}
S(x,y,z) &=&
I(x,y,z) + s \left [ \frac{x}{2} I(x'',y,z)-\frac{1}{8} \right ]
+ s^2 \left [ 
\frac{x}{6} I(x''',y,z) + \frac{x^2}{12} I(x'''',y,z)
\right] + \ldots
,
\\
S(0,x,x) &=&
I(0,x,x) + s \left [ -(\lnbarx)/2-1/8 \right ] + s^2/36 x^2 + \ldots
.
\end{eqnarray}
Taking the derivative with respect to $x$ yields
\begin{eqnarray}
T(x,y,z) &=& -I(x',y,z) 
+ s \left [
-\frac{1}{2} I(x'',y,z) -\frac{x}{2} I(x''',y,z) 
\right ]
\nonumber \\ &&
+ s^2 \left [ 
-\frac{1}{6} I(x''',y,z) - \frac{x}{3} I(x'''',y,z)
-\frac{x^2}{12} I(x''''',y,z) \right ]
+ \ldots
,
\\
T(x,0,x) &=&
(1-\lnbarx )^2/2 + s/4x + s^2/72 x^2 + \ldots
.
\end{eqnarray}
The infrared-safe $\Tbar$ function has the expansion
\begin{eqnarray}
\Tbar (0,x,y) &=&
\frac{1}{(x-y)^2} \left [
(x + y) I(0, x, y) + 2 A(x) A(y) -2 x A(x) - 2 y A(y) + (x+y)^2
\right ]
\nonumber \\ &&
+ \frac{s}{2(x-y)^4}
\left [ 4 x y I(0, x, y) + (x+y) \lbrace
2 A(x) A(y) + (x-3y) A(x) + (y-3 x) A(y) + 4 x y \rbrace \right ]
\nonumber \\ &&
+ \frac{s^2}{12(x-y)^6}
\bigl [24 x y (x + y) I(0,x,y) + 12 (x+y)^2 A(x) A(y) 
+(2 x^3 + 20 x^2 y - 42 x y^2 - 28 y^3) A(x)  
\nonumber \\ &&
+(2 y^3 + 20 y^2 x - 42 y x^2 - 28 x^3) A(y)
-3 (x^4 + y^4) + 8x y (x^2 + y^2) + 86 x^2 y^2 \bigr ]
+ \ldots
,
\\   
\Tbar (0,x,x) &=& -\frac{1}{2} \lnbar^2 x -\lnbarx -\frac{3}{2}
+ \frac{s}{36 x} \left [6 \lnbarx +1 \right ] 
+ \frac{s^2}{900 x^2} \left [ 15 \lnbarx - 19 \right ]
+ \ldots
.
\end{eqnarray}
For the $U$ integral,
\begin{eqnarray}
U(x,y,z,u) &=& 
\frac{1}{y-x} \left [I(x,z,u) - I(y,z,u) \right ] +
s \Bigl [
\frac{x}{(y-x)^3} \left (I(x,z,u) - I(y,z,u) \right )
+ \frac{x}{(y-x)^2} I(x',z,u) 
\nonumber \\ &&
+ \frac{x}{2(y-x)} I(x'',z,u) \Bigr ]
+ s^2 \Bigl [
\frac{x (x+y)}{(y-x)^5} \left (I(x,z,u) - I(y,z,u) \right )
+ \frac{x (x+y)}{(y-x)^4} I(x',z,u)
\nonumber \\ &&
+ \frac{x (x+y)}{2(y-x)^3} I(x'',z,u)
+ \frac{x (x+y)}{6(y-x)^2} I(x''',z,u)
+ \frac{x^2}{12(y-x)} I(x'''',z,u)
\Bigr ] + \ldots
,
\\
U(x,x,z,u) &=& - I(x',z,u) +s \left [-\frac{x}{6} I(x''',z,u) \right ]
+ s^2 \left [ -\frac{x}{24} I(x'''',z,u) -\frac{x^2}{60} I(x''''',z,u) 
\right ] + \ldots
.
\end{eqnarray}
For the master integral,
\begin{eqnarray}
M(x,y,z,u,v) &=&
\frac{1}{(x-z)(y-u)} \left [
I(x,y,v) -I(x,u,v) -I(z,y,v) +I(z,u,v) 
\right ]
+\frac{s}{4(y-u)^2(x-z)^2} \Bigl [ 
\nonumber \\
&&
\Bigl \lbrace
\bigl [4u+4z-x-y-2v +\frac{4x(y-u)}{x-z} +\frac{4y(x-z)}{y-u}
\bigr ] [I(x,y,v) - 
I(x,u,v)]
+ (u-y) I(x,u,v)
\nonumber \\
&&
+ v (x + y -v) (x + y) I(x,y,v'')
+ v (u v + v x- 2 x y - 2 u z + 2 y z -u^2-x^2) I(x,u,v'')
\nonumber \\
&&
+ 2 x y (x + y) I(x',y',v)
- 2 x u (x + u) I(x',u',v)
+ x (v-x - 3 y) I(x',y,v) 
\nonumber \\
&&
+ y (v-y - 3 x) I(x,y',v)
+ x (x + 4 y -u-v) I(x',u,v)
+ u (u + 4 z -x-v) I(x,u',v)
\Bigr \rbrace
\nonumber \\
&&
+ \Bigl \lbrace 
(x,y) \leftrightarrow (z,u)
\Bigr \rbrace
\Bigr ] + \ldots
,
\\
M(x,y,x,u,v) &=& 
\frac{1}{y-u} \left [I(x',y,v) - I(x',u,v) \right ] 
+ s \Bigl [
-\frac{u+y}{24 x y u}
+\frac{u+y}{2(y-u)^3} [I(x',y,v)- I(x',u,v)]
\nonumber \\ &&
-\frac{1}{2(y-u)^2} [u I(x',u',v) + y I(x',y',v)]
+\frac{(u+y)}{24 y u (y-u)} [
u(x+u-v) I(x'',u',v) 
\nonumber \\ &&
- y (x+y-v) I(x'',y',v) 
+(v+2u-x) I(x'',u,v) - (v+2y-x) I(x'',y,v)]  
\nonumber \\ &&
-\frac{x}{12 y u } [u I(x''',u,v) + y I(x''',y,v)]
\Bigr ] + \ldots
,
\\
M(x,y,x,y,v) &=& I(x',y',v) 
\nonumber \\ &&
+ \frac{s}{24 x y} \Bigl [
5 + 6 v I(v'',x,y) +2 v (4 v-x-y) I(v''',x,y) 
+v^2 (v-x-y) I(v'''',x,y) \Bigr ] + \ldots
.
\end{eqnarray}
In theories with massless vector bosons, special cases like 
$M(x,y,x,0,x)$
can arise, in which denominators implicit in the previous expressions 
threaten to
vanish. 
However, those cases are easily obtained from the preceding, by noting 
that
e.g.~$u I(x,y,u')$ vanishes as $u \rightarrow 0$, since $I(x,y,u')$
diverges only logarithmically in that limit.

\section{Analytical results\label{analytical}}
\setcounter{equation}{0}

As noted in the Introduction, for favorable mass and momentum 
configurations the basis integrals can be, and in many cases have been
\cite{Rosner}-\cite{Martin:2001vx},
computed analytically.
The results for $s=0$ were given in the previous section. I will not
consider other special values of $s$ in this section; they do not 
typically
arise in mass-independent (as opposed to on-shell) renormalization 
schemes. 
The remaining cases involve
vanishing squared masses, which arise in theories with unbroken gauge
symmetries, and as approximations to theories with large mass hierarchies.
Results for these cases can be obtained by analytically integrating the
differential equations presented in section \ref{diffeqs}, with the 
initial conditions of section \ref{exps}, taking due care with the
branch cuts.
In this section, I will review results obtained in this manner, most of
which have already been derived by dispersion relation and other methods.

To compactify the notation, define the quantities 
\begin{eqnarray}
t_{abc} = \frac{a+b-c + \Delta_{abc}^{1/2}}{2a},
&\qquad\qquad&
r_{abc} = \frac{a+b-c - \Delta_{abc}^{1/2}}{2a}.
\end{eqnarray}
They obey
\begin{eqnarray}
t_{abc} = \frac{1}{1-t_{bca}} = 1-\frac{1}{t_{cab}} = 
1 - r_{acb} = \frac{1}{r_{bac}} = \frac{r_{cba}}{{r_{cba}-1}}.
\end{eqnarray}
These are exactly the changes of variables that occur in dilogarithm 
functional identities \cite{Lewin},
making the presentation of formulas below highly non-unique.
To resolve branch cuts in the following consistent with the standard 
conventions for polylogarithms \cite{Lewin}, it is crucial that $s$
is always given an infinitesimal positive imaginary part.

For the one-loop formulas, the well-known result is:
\begin{eqnarray}
B(x,y) &=& 
2 - r_{sxy} \lnbarx - t_{syx} \lnbary +
(\Delta^{1/2}_{sxy}/s) \ln(t_{xys}) ,
\\
B(0,x) &=& 2 - \lnbarx + (x/s-1) {\rm ln}(1-s/x) 
\label{B0xanal} ,
\\
B(0,0) &=& 2 - \lnbar(-s) .
\end{eqnarray}

The two-loop vacuum integral is given by 
\cite{vanderBij:1983bw,Ford:hw,
Ford:pn,Davydychev:1992mt,Berends:1994ed,Caffo:1998du,Martin:2001vx}
\begin{eqnarray}
I(x,y,z) &=& 
\frac{1}{2} \left [(x-y-z) \lnbary \lnbarz
+(y-z-x) \lnbarx \lnbarz
+(z-x-y) \lnbarx \lnbary \right ]
+ 2 (x \lnbarx +y \lnbary +z \lnbarz)
\nonumber \\ &&
-\frac{5}{2}(x+y+z)
+ \Delta_{xyz}^{1/2} \left [
\dilog(r_{xyz}) + \dilog(r_{xzy}) - \ln(r_{xyz})\ln(r_{xzy}) +
\frac{1}{2} \ln(y/x) \ln(z/x) - \zeta(2) \right ] 
\end{eqnarray}
when $x>y,z$, and otherwise by the appropriate symmetry permutation
of the arguments. Some special limits are
\begin{eqnarray}
I(0,x,y) &=& (x-y) \left [ \dilog(y/x) - \lnbar(x-y) \ln(x/y) + 
(\lnbarx)^2/2
- \zeta(2) \right ] 
%\nonumber && \\
+x \lnbarx(2- \lnbary) + 2 y \lnbary -5 (x+y)/2 ,
\phantom{x}
\\
I(0,x,x) &=& x [-\lnbar^2 x + 4 \lnbarx -5], \\
I(0,0,x) &=& x [-(\lnbarx)^2/2 + 2 \lnbarx -5/2 -\zeta(2)] .
\end{eqnarray}

When the masses are all very small, the two-loop basis integrals 
defined in this paper are
\begin{eqnarray}
S(0,0,0) &=& \frac{13s}{8} - \frac{s}{2} \lnbar(-s) ,\\
\Tbar (0,0,0) &=& -\frac{1}{2}\left [\lnbar(-s)-1 \right ]^2 ,\\
U(0,0,0,0) &=& \frac{1}{2} \left [\lnbar(-s)- 3 \right ]^2 +1 ,\\
M(0,0,0,0,0) &=& -6 \zeta (3)/s .
\label{Mlarges}
\end{eqnarray}
This should provide a useful quick comparison between other conventions
and the ones used here.

For the $S$ and $T$ functions with one vanishing mass and the others
arbitrary, one finds \cite{Berends:1994ed}:
\begin{eqnarray}
S(0,x,y) &=& 
(y-x) [\dilog (t_{xsy}) + \dilog (r_{xsy})] 
-y (1-x/s) {\rm ln}(t_{xys}) {\rm ln}(r_{xys}) 
+[(x+y+s) \Delta^{1/2}_{sxy}/4s] [{\rm ln}(t_{xys}) -{\rm ln}(r_{xys})] 
\nonumber \\ && 
+ (y-x) [\lnbarx]^2/2 -y \lnbarx \lnbary 
+(2 x-s/4) \lnbarx  + (2 y -s/4) \lnbary
+ [(y^2 - x^2)/4s] {\rm ln}(x/y) 
\nonumber \\ && 
-2x-2y + 13 s/8 ,
\\
T(x,0,y) &=& \dilog (t_{xsy}) + \dilog (r_{xsy}) 
+\ln(r_{xys}) [ y \ln(r_{yxs}) + \Delta^{1/2}_{sxy} ]/s
+r_{syx} \ln(y/x) +\frac{1}{2}[\lnbarx-1]^2 -1 .
\end{eqnarray}
Here I have deliberately chosen a presentation that does not make
manifest\footnote{Of course, the manifest symmetry under
$x\leftrightarrow y$ can be restored using dilogarithm identities.}
the symmetry under $x\leftrightarrow y$. This makes the
formulas slightly smaller, and also eases the taking of the limit
$y \rightarrow 0$: 
\begin{eqnarray}
S(0,0,x) &=& -x \dilog (s/x) -x (\lnbarx)^2/2 + (2 x -s/2) \lnbarx +
[(x^2 - s^2)/2 s] \ln(1 - s/x) + 13 s/8 - [2  + \zeta (2)] x ,
\phantom{xx}
\label{S00xanal}
\\
T(x,0,0) &=& \dilog (s/x) + (\lnbarx)^2/2 - \lnbarx 
+ (1-x/s) \ln(1 - s/x) -1/2 + \zeta(2) .
\label{Tx00anal}
\end{eqnarray}

The analytical expression for the $\Tbar$ integral evidently cannot be
obtained from those for $S,T$. By integrating the differential equation
(\ref{dTbards}), I find
\begin{eqnarray}
\Tbar (0,x,y) &=&
(1-2 t_{sxy}) \dilog (t_{xsy}) +(1-2 r_{sxy}) \dilog (r_{xsy})
+ (2 \Delta_{sxy}^{1/2}/s) \dilog (-x r_{xys}/\Delta_{sxy}^{1/2})
\nonumber \\ &&
+ \frac{\Delta_{sxy}^{1/2}}{s} \Bigl [
\lbrace \ln (x t_{xys}/\Delta_{sxy}^{1/2}) +2 \ln(r_{xys}) \rbrace^2 
+ (1 - \lnbary)\ln(r_{xys})
+2 \ln(y/x) \ln(\Delta_{sxy}^{1/2}/x) 
\nonumber \\ &&
+\lbrace 5 \lnbarx\lnbary -3 \lnbar^2 x- 2 \lnbar^2 y + \ln(x/y)\rbrace /2
+ 2 \zeta(2) \Bigr ]
+ [(s-x-2 \Delta_{sxy}^{1/2})/s] \ln^2(r_{xys}) 
\nonumber \\ &&
+ (1-x/s) \ln(x/y) \ln(r_{xys}) 
+ (1-\lnbarx) [(x/s-y/s) \ln(x/y) -(1-\lnbary)]/2 
,
\\ 
\Tbar (0,0,x) &=& -\dilog (s/x) - (\lnbarx)^2/2 + \lnbarx 
+ (1-x/s) \ln(1 - s/x) \lbrace 1 - \lnbarx - {\rm ln}(1 - s/x) \rbrace 
-1/2 - \zeta(2)
.
\label{Tbar00xanal}
\end{eqnarray}

Useful cases for the $U$ and $V$ integrals that arise in 
unbroken gauge theories are compactly written in terms of
the preceding integrals:
\begin{eqnarray}
U(x,y,y,0) &=&
-T(y,0,x) + (2 - \lnbary) B(x,y) +1 ,
\label{Uxyy0anal}
%\\ 
%&=&
%\dilog(t_{xsy}) + \dilog (r_{xsy}) 
%+ (1-x/s) {\rm ln}(r_{xys}) {\rm ln}(r_{yxs})
%+(\lnbary-3)[(t_{sxy}-1/2){\rm ln}(x/y)
%+(\Delta^{1/2}_{sxy}/s){\rm ln}(r_{xys})]
%\nonumber
%\\ &&
%+ (\lnbar^2 x +\lnbar^2 y -\lnbarx \lnbary -3 \lnbarx -3 \lnbary +11)/2
%\end{eqnarray}
%\begin{eqnarray}
\\
V(x,y,y,0) &=&
\frac{1}{2y} \left [
\Tbar (0,x,y) -T(y,0,x) - \lnbary B(x,y) \right ]
+ (\lnbary-2) B(x,y') .
\end{eqnarray}
The last integral was obtained using equation
(\ref{dUdyy0}) and the
definition (\ref{defV}).
Equivalent results were found in \cite{Djouadi:1987di}. 

Some other special limits of the $U$ integral that can be quickly obtained
using the differential equation method are:
\begin{eqnarray}
U(x,0,0,0) &=& \dilog (s/x) + (1-x/s) {\rm ln}(1 - s/x) [
\lnbarx -3 +{\rm ln}(1 - s/x)] + (\lnbarx)^2/2 - 3 \lnbarx + 11/2 + 
\zeta(2) ,
\\ 
U(0,x,0,0) &=& (1-x/s) \lbrace 
\dilog(s/x) + [\lnbar(-s)-2] \ln(1-s/x) \rbrace
- \lnbar(-s) +(\lnbarx-2)^2/2 + 7/2 + \zeta(2)
, \\
U(0,0,0,x) &=& -(1+x/s) \dilog (s/x) -(\lnbarx)^2/2 - 2 \lnbarx + 
(\lnbarx -1)\lnbar(-s) -2 (1-x/s) {\rm ln}(1 - s/x) 
\nonumber \\ &&
+ 11/2 - \zeta(2) .
\end{eqnarray}
Equivalent results were obtained in \cite{Scharf:1993ds}.

By integrating the differential equation (\ref{dUds}) with the first 
argument vanishing, I find:
\begin{eqnarray}
U(0,z,x,y) &=&
 (y-x) [ \dilog (t_{xsy}) + \dilog (r_{xsy})]/z 
-(y/s) {\rm ln}(t_{xys}) {\rm ln}(r_{xys}) 
+ (1-z/s) \Bigl \lbrace
\nonumber \\ &&
(t_{zxy}-1/2) \Bigl [
\dilog(1-t_{xys}r_{yxz}) + \dilog(1-r_{xys}r_{yxz})  
-\dilog(t_{yzx}) -\dilog(t_{xzy}) -\ln(t_{xyz}) \ln(1-s/z)
\nonumber \\ &&
-\eta(t_{xyz},r_{yxs})\ln(1-t_{xys}r_{yxz})
-\eta(t_{xyz},t_{yxs})\ln(1-r_{xys}r_{yxz})
-\eta(t_{xyz},1/t_{xyz}) [\ln(t_{yzx})+\ln(t_{xzy}) ]
\Bigr ]
\nonumber \\ &&
+(r_{zxy}-1/2) \Bigl [ \dilog(1-t_{xys}t_{yxz}) +
\dilog(1-r_{xys}t_{yxz})
-\dilog(r_{yzx}) -\dilog(r_{xzy}) -\ln(r_{xyz}) \ln(1-s/z)
\nonumber \\ &&
-\eta(r_{xyz},r_{yxs})\ln(1-t_{xys}t_{yxz})
-\eta(r_{xyz},t_{yxs})\ln(1-r_{xys}t_{yxz})
-\eta(r_{xyz},1/r_{xyz}) [\ln(r_{yzx})+\ln(r_{xzy}) ]
\Bigr ]
\nonumber \\ &&
+ (\lnbarx+\lnbary-4) \ln(1-s/z)/2 
+[\ln(r_{xys})-1]^2/4  +[\ln(t_{xys})-1]^2/4 -\ln^2(x/y)/4 - \ln(x/y)/2
\Bigr \rbrace
\nonumber \\ &&
-I(x,y,z)/z + (\Delta_{sxy}^{1/2}/2s) [\ln(t_{xys}) - \ln(r_{xys})]
-(y/z) \lnbarx\lnbary + (2x/z-1/2) \lnbarx + (2y/z-1/2) \lnbary
\nonumber \\ &&
+[(y-x)/2s] \ln(x/y) + 5 (z-x-y)/2z + z/2s
+[(y-x)/2 z] \lnbar^2 x ,
\label{U0zxyanal}
\end{eqnarray}
where the function
\begin{eqnarray}
\eta(a,b) = \ln(ab) - \ln(a) - \ln(b)
\end{eqnarray}
is employed to properly treat the branch cuts. As far as I know, this is
the first analytical computation of a two-loop self-energy diagram with 
generic 
$s$ and three distinct non-zero masses. I have checked it numerically
using the method of the next section. 

Broadhurst has computed \cite{Broadhurst:1987ei} the master integral for
the special limits needed in unbroken gauge theories:
\begin{eqnarray}
M(x,x,y,y,0) &=& \left [ F_3^+(t_{xys}) + F_3^+(t_{yxs})
-4F_3^+(\sqrt{x/y}\, t_{xys}) -4 F_3^-(\sqrt{x/y}\, t_{xys}) - 6 \zeta(3)
\right ]/s ,
\\
M(x,x,0,0,0) &=& \left [ F_3^+ (x/(x-s))- 6 \zeta(3) \right ]/s, 
\end{eqnarray}
where
\begin{eqnarray}
F_3^+ (z) &=& 
6 \trilog (z) - 4 {\rm ln}(z) \dilog(z) -{\rm ln}(1-z) {\rm ln}^2 (z)
, \\
F_3^-(z) &=& 6 \trilog (-z) - 4 {\rm ln}(z) \dilog(-z) 
-{\rm ln}(1+z){\rm ln}^2(z) .
\end{eqnarray}
Another special case is \cite{Broadhurst:1987ei}
\begin{eqnarray}
M(x,0,0,0,0) &=& 
\left [ F_3^+ (x/(x-s))-F_3^+ (s/(s-x)) - 6 \zeta(3) \right ]/2s .
\end{eqnarray}
I have checked that these results are satisfied by the differential
equation (\ref{dMds}), using the other analytical results above.
(Straightforward integration of equation (\ref{dMds}) provides 
more complicated expressions, not given here, which are then 
evidently related
to the above by some trilogarithm identities. The equivalence was checked
numerically.)

Some other special cases that have been computed in the literature will be
omitted here for brevity. Ref.~\cite{Broadhurst:1987ei} also found
$M(x,0,x,0,x)$, while ref.~\cite{Scharf:1993ds}
obtained the equivalent of $U(x,y,0,0)$ and $M(x,0,y,0,0)$ and
$M(0,0,0,0,x)$, and ref.~\cite{Fleischer:1998nb} has $M(x,0,0,x,0)$,
$M(x,0,0,0,x)$, $M(x,x,x,0,0)$, and $M(x,x,x,x,0)$.

\end{widetext}

\section{Numerical evaluation by differential equations\label{numerical}}
\setcounter{equation}{0}

A method for using the differential equations in $s$ to 
numerically compute basis integrals has been formulated 
by Caffo, Czyz, Laporta, and Remiddi in 
\cite{Caffo:1998du}-\cite{Caffo:2002wm}. 
We can now apply the same strategy to compute
the values of all of the basis integrals, using the differential equations
worked out in section \ref{diffeqs}. 

Consider a master integral $M(x,y,z,u,v)$ that occurs in a self-energy
function. Typically, one will also need some or all of the basis integrals
that arise from removing one or more propagators. These can all be 
obtained simultaneously by solving the 
system of coupled first-order ordinary differential equations in the 15 
dependent quantities
\begin{eqnarray}
&& M(x,y,z,u,v),\> U(x,z,u,v),\> U(y,u,z,v),\>
\nonumber \\ &&
U(z,x,y,v),\> U(u,y,x,v),\>
T(x,u,v),\> 
T(y,z,v),\>
\nonumber \\ &&
T(z,y,v),\> T(u,x,v),\>
T(v,x,u),\>  T(v,y,z),\>
\nonumber \\ &&
S(x,u,v),\> S(y,z,v),\>
B(x,z),\> B(y,u) ,
\end{eqnarray}
with $x,y,z,u,v$ fixed and $s$ as the independent variable. 
The relevant differential equations in addition to (\ref{dMds}) are
(\ref{dBds}), 
(\ref{dSds}),
(\ref{dTds}), (\ref{dUds}), and others obtained by
obvious permutations.
Since the $B$ functions are
known analytically, one need not treat them as among the dependent
variables, but it is probably more economical in terms of
computer processing time to do so. Other than the term involving $B(x,z) 
B(y,u)$ in the differential equation for the master integral 
$M(x,y,z,u,v)$, the system
of equations is linear.

Standard computer numerical methods (for example, Runge-Kutta, or
improvements thereof) are used to evolve the differential equations from
$s=0$ to the desired $s$. Since the physical-sheet $s$ is always taken to
have an infinitesimal real imaginary part, and branch cuts lie along the
real $s$ axis, one should take the contour of integration to lie in the
upper-half complex plane. Reference \cite{Caffo:2002wm} suggests using a
rectangular contour going from $0$ to $i h$ to $s+ih$ to $s + i
\varepsilon$, where $h$ is chosen large enough to stay away from
singularities on the real $s$ axis. Independence of the choice of $h$, and
more generally on the choice of contour in the upper half-plane, provides
a useful check on the numerical convergence.

At the start of the contour at $s=0$, the appearance of $s$ on the
left-hand sides of the differential equations
requires that the initial data for derivatives of the basis functions with
respect to $s$ are provided, along with the initial values.
(Alternatively, one can start the running at a point very slightly
displaced from $s=0$.) These are obtained from the expansions in section
\ref{exps}. I find that it is often better to run $s M(x,y,z,u,v)$
rather than the master integral itself. The method is always very fast and
arbitrarily accurate, except sometimes when $s$ is equal or extremely 
close to one of the thresholds where the denominators in the 
differential equations vanish. Even these cases can be efficiently
computed without performing special analytical expansions around the 
thresholds, as will be explained below.

I have implemented this method in a computer program in order to test the
method, and for use in future applications. When 
doing so, it is useful to note that all quantities other than
$s$ remain constant in the course of a Runge-Kutta routine. Therefore,
although the coefficients of various powers of $s$ in the numerators and
denominators of the coefficient functions are mildly complicated functions
of $x,y,z,u,v$, they only need to be computed once. Comparison with 
specific numerical examples for the master integral in
ref.~\cite{Bauberger:1994hx} and the sunrise integrals in
\cite{Caffo:2002ch} yields agreement. Note that the first of
these comparisons is actually a test of the equations and the method for
all of the basis functions, not just the master integral $M$, since any 
error in any of the basis functions would feed into a discrepancy for the 
master integral.

\begin{figure}[tb]
\includegraphics[width=8.0cm]{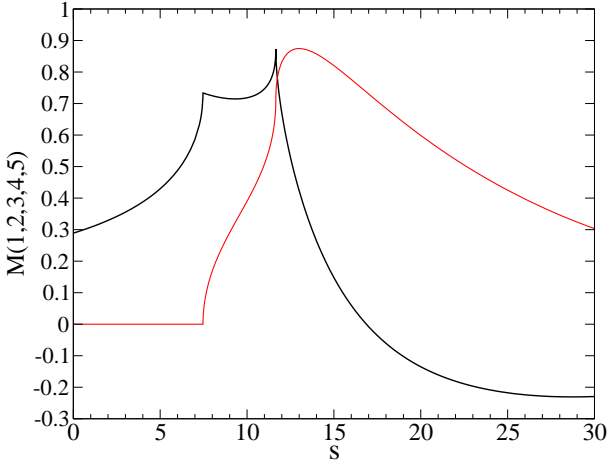}
\caption{\label{fig:M12345}The master 
integral $M(1,2,3,4,5)$, as a 
function of $s$. The heavier line is the real part, and the lighter is 
the imaginary part.}
\end{figure}

As an example, I consider the master integral and its subordinates for
the case $Q=1$, $x=1$, $y=2$, $z=3$, $u=4$, and $v=5$.
The result for the master integral $M(1,2,3,4,5)$ as a function of $s$ is 
shown in
Figure \ref{fig:M12345}. 
Although the dependence on $s$ near the two-particle thresholds 
$s=(1+\sqrt{3})^2 \approx 7.464$ and $(\sqrt{2}+2)^2 \approx 11.657$ 
is sharp, these points [and the three-particle thresholds 
$(1+2+\sqrt{5})^2 \approx 27.416$ and $(\sqrt{2}+\sqrt{3}+\sqrt{5})^2 
\approx 28.970$]
did not present any numerical 
problems. The value of the master integral at $s=0$ is 
$[I(1,2,5)-I(1,4,5)-I(2,3,5)+I(3,4,5)]/4 \approx 0.28889224$.
The asymptotic limit in which equation~(\ref{Mlarges}) is reasonably 
accurate is very far to the right of the end of the graph.
Values for all of the basis integrals at 
$s=10$ found from the simultaneous numerical solution to the
differential equations are:
\begin{eqnarray}
M(1,2,3,4, 5)   &=& \phantom{-}0.71833535 + 0.39016220 i 
\nonumber
\\
\nonumber
U(1, 3, 4, 5) &=& -4.85695306 - 2.12756034 i \\
\nonumber
U(3, 1, 2, 5) &=& -3.99263620 - 1.79951451 i \\
\nonumber
U(2, 4, 3, 5) &=& -3.08641797 \\
\nonumber
U(4, 2, 1, 5) &=& -2.23235894 \\
\nonumber
S(1, 4, 5) &=&    -9.56660679 \\
\nonumber
T(4, 1, 5) &=&    -0.03036018 \\
\nonumber
T(5, 1, 4) &=&     \phantom{-}0.51591658 \\
\nonumber
T(1, 4, 5) &=&    -3.01221172 \\
\nonumber
S(5, 2, 3) &=&    -7.67047979 \\
\nonumber
T(5, 2, 3) &=&     \phantom{-}0.44677524 \\
\nonumber
T(2, 3, 5) &=&    -1.69451693 \\
\nonumber
T(3, 2, 5) &=&    -0.78612788 \\
\nonumber
B(1,3) &=&       \phantom{-}  0.77930384 + 1.53905980 i \phantom{xxx}\\
B(2,4) &=&        -0.05151328 \> .
\end{eqnarray}

\begin{figure}[tb]
\includegraphics[width=8.0cm]{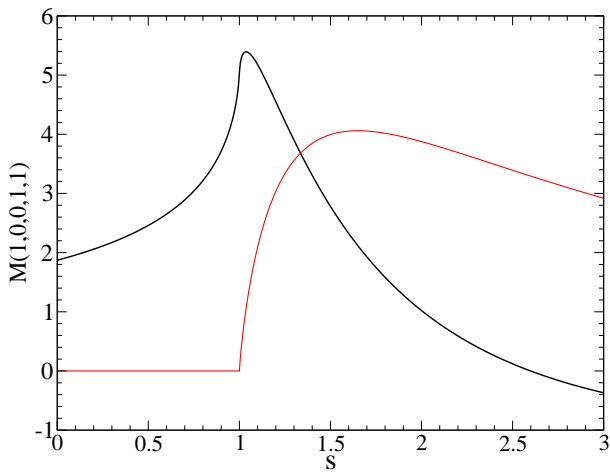}
\caption{\label{fig:M10011}The master 
integral $M(1,0,0,1,1)$, as a 
function of $s$. The heavier line is the real part, and the lighter is 
the imaginary part.}
\end{figure}

As a second test case, consider the master integral $M(x,0,0,x,x)$,
which occurs in QED and QCD. This case does not satisfy the
criterion for solvability in terms of generalized polylogarithms
mentioned in the introduction, but a simple integral representation
has been worked out in
\cite{Broadhurst:1987ei}. Following the method adopted here,
the full system of differential equations simplifies to:
\begin{eqnarray}
s \frac{d}{ds} B(0,x) &=& \frac{x}{s-x} \left [ B(0,x) +\lnbarx 
- 1-s/x \right ]\! ,
\phantom{xxx}
\\
s \frac{d}{ds} S (0,0,x) &=& x T(x,0,0) + S(0,0,x) - x \lnbarx 
\nonumber \\ && 
+ 2 x -s/2,
\\
s \frac{d}{ds} S (x,x,x) &=& 3x T(x,x,x) + S(x,x,x) - 3x\lnbarx 
\nonumber \\ && 
+ 6 x -s/2,
\end{eqnarray}
\begin{widetext}
\begin{eqnarray}
s \frac{d}{ds} \Tbar (0,0,x) &=& \frac{1}{(s-x)^2} \bigl [
x (s-x) \Tbar(0,0,x) + x (3s+x) T(x,0,0) + 2 (s+x) S(0,0,x)
+ x (s-3x) \lnbarx 
\nonumber \\ && 
+ 5 x^2 + 3 s x/4 - 9 s^2/4 \bigr ] ,
\\
s \frac{d}{ds} T (x,0,0) &=& \frac{1}{s-x} \bigl [
2x T(x,0,0) + 2 S(0,0,x) + (s-2x) \lnbarx + 4x-9s/4 \bigr ]
, \\
s \frac{d}{ds} T(x,x,x) &=& \frac{1}{(s-x)(s-9x)} \bigl [
2x (5s-9x) T(x,x,x) + 2  (s-3x) S(x,x,x) -2sx \lnbar^2 x 
\nonumber \\ &&
+ (s^2 - 5 s x + 18 x^2)\lnbarx 
-36 x^2 + 67 s x/4 - 9 s^2/4 \bigr ]
, \\
s \frac{d}{ds} U(x,0,x,x) &=& \frac{1}{(s-x)^2} \bigl [
x (s-x) U(x,0,x,x) - x (5s+3x) T(x,x,x) - 2 (s+x) S(x,x,x) 
\nonumber \\ &&
+ (s+x) x (7 \lnbarx - \lnbar^2 x -11) +s(s+x)/4 \bigr ] 
, \\
s \frac{d}{ds} U(0,x,x,0) &=& \frac{1}{s-x} \bigl [
x U(0,x,x,0) - x T(x,0,0) - 2 S(0,0,x) 
+ x (5 \lnbarx - \lnbar^2 x -7) +s/4 \bigr ] 
, \\
\frac{d}{ds} [sM(x,0,0,x,x)] &=& \frac{1}{(s-x)^2}
\bigl [ (s+3x) T(x,x,x) + (s+x) T(x,0,0) + 2 S(x,x,x) +
2 S(0,0,x) + s B(0,x)^2 
\nonumber \\ &&
+ (2 \lnbarx -4) s B(0,x) +
2 x (\lnbarx -3)^2 -3s/2 \bigr ] .
\end{eqnarray}
\end{widetext}
The value of the master integral obtained for $x=1$ and as a function
of $s$ is shown in Figure \ref{fig:M10011}. 
In this example, it turns out that there are numerical
problems, but only extremely close to the double threshold at $s=x=1$, 
where it is known that
\cite{Broadhurst:1987ei}
\begin{eqnarray}
M(1,0,0,1,1) =  \pi^2 \ln2 -3 \zeta(3)/2 \approx 
5.03800311.\phantom{xx}
\label{M10011anal}
\end{eqnarray}
In mass-independent renormalization schemes, this is not an issue since
the tree-level mass appearing as the argument of the function is not
exactly the same as the pole mass where one will need to evaluate the 
self-energy. In on-shell type schemes, one could 
find the threshold value analytically, but one can also use the 
following general procedure to find threshold values with high accuracy.
Near each threshold $s_0$, the loop functions have expansions of the form
\begin{eqnarray}
F(s) &=& F(s_0) 
+ r [a_1 + b_1 \ln r + c_1 \ln^2 r]
\nonumber \\ &&
+ r^2 [a_2 + b_2 \ln r + c_2 \ln^2 r ] 
+ \ldots
\end{eqnarray}
where $r = 1-s/s_0$.
Now one can use the Runge-Kutta method to evaluate the loop functions
at, say, several points $s=s_0 \pm n\delta$ (for small integers $n$), 
and then simply solve for 
the coefficients in the expansion, in particular $F(s_0)$.
In the present example, I find that choosing $\delta = 10^{-4}$, 
where there are definitely no numerical problems, 
and $n=1,2,3,4$ is good enough to obtain the threshold 
values for $s=x=Q=1$ 
to better than 9 significant digits. The results are:
\begin{eqnarray}
B(0,1) &=& \phantom{-}       2.00000000 
\nonumber 
\\
S(0,0,1) &=&                -3.66486813 
\nonumber
\\
T(1,0,0) &=& \phantom{-}     2.78986813 
\nonumber
\\
\Tbar(0,0,1) &=&            -3.78986813 
\nonumber
\\
U(0,1,1,0) &=& \phantom{-}   2.21013187 
\nonumber
\\
M(1,0,0,1,1) &=& \phantom{-} 5.03800311
\nonumber
\\
S(1,1,1) &=&                -4.37500000 
\nonumber
\\
T(1,1,1) &=&                -0.50000000 
\nonumber
\\
U(1,0,1,1) &=&              -1.07973627
\> .
\end{eqnarray}
Of these, the first six are checked using the analytic formulas
(\ref{B0xanal}), (\ref{S00xanal}), (\ref{Tx00anal}), (\ref{Tbar00xanal}),
(\ref{Uxyy0anal}),
and
(\ref{M10011anal}), while the next two can now be seen ``experimentally" 
to
have the analytical values $S(1,1,1) = -35/8$ and $T(1,1,1) = -1/2$ 
at threshold.

To be extra-safe, a computer code can be configured to always trap the
threshold and pseudo-threshold cases for evaluation in this manner.
This is easy to do in an automated way,
since the potentially dangerous points are always known in advance as the 
roots of the denominators of the differential equations or from inspection 
of the Feynman diagrams. 

\section{Outlook}
%"Stupidity is reaching a conclusion."

In this paper I have studied the properties of a minimal basis of integral
functions for two-loop self energies. These results include a complete set 
of formulas allowing for their automated numerical computation using
differential equations, following the same strategy as was put forward in 
\cite{Caffo:1998du}-\cite{Caffo:2002wm}. 
It might be useful to review some of the
advantages of this method:
\begin{itemize}
\item The basis integrals can be computed for any values of all 
masses and $s$, to arbitrary accuracy.
\item All of the necessary basis integrals are obtained 
simultaneously in a single numerical computation.
\item Branch cuts are automatically dealt with correctly by 
choosing a contour in the upper-half complex $s$ plane.
\item Simple checks on the numerical accuracy follow from changing 
the choice of contour.
\end{itemize}
The Tarasov recurrence relation algorithm 
\cite{Tarasov:1997kx,Mertig:1998vk}
can be used
to reduce any two-loop self-energy to linear combinations of these 
functions, with coefficients depending on the masses and couplings of the
theory. Recently, I have used this basis and the methods of 
computation described here to obtain the leading 
two-loop momentum-dependent corrections to the neutral Higgs boson masses in
minimal supersymmetry in a mass-independent renormalization scheme. That 
result will appear soon.

\vspace{0.25in}

This work was supported by the National Science Foundation under Grant No.
0140129.

\end{document}